\documentclass[12pt,a4j]{article}
\usepackage[dvips]{graphicx}
\usepackage{enumerate}
\usepackage{amsmath}
\usepackage{amsfonts}

\begin{document}
\baselineskip=7mm
\centerline{\bf A direct method of solution for the Fokas-Lenells derivative }\par
\centerline{\bf nonlinear Schr\"odinger equation: I. Bright soliton solutions}\par
\bigskip
\centerline{Yoshimasa Matsuno\footnote{{\it E-mail address}: matsuno@yamaguchi-u.ac.jp}}\par

\centerline{\it Division of Applied Mathematical Science,}\par
\centerline{\it Graduate School of Science and Engineering} \par
\centerline{\it Yamaguchi University, Ube, Yamaguchi 755-8611, Japan} \par
\bigskip
\bigskip
\leftline{\bf Abstract}\par
\noindent We develop  a direct method of solution for finding the bright $N$-soliton solution of the Fokas-Lenells derivative
nonlinear Schr\"odinger equation. The construction of the solution  is performed by means of a purely algebraic
procedure using an elementary theory of determinants and does not rely on the inverse scattering transform method.
We present  two different expressions of the solution  both of which are expressed as a ratio of determinants.
We then investigate the properties of the solutions and find several new features. 
Specifically, we derive the formula for the phase shift caused by the collisions of bright solitons.

  \par

\newpage
\leftline{\bf  1. Introduction} \par
\bigskip
\noindent In this study, we consider the following Fokas-Lenells (FL) derivative nonlinear Schr\"odinger (DNLS) equation:
$$u_{xt}=u-2{\rm i}|u|^2u_x, \eqno(1.1)$$
where $u=u(x,t)$ is a  complex-valued function of $x$ and $t$, 
and subscripts $x$ and $t$ appended to $u$ denote  partial differentiations.  The original version of the FL DNLS equation has been derived as an
integrable generalization of the NLS equation using  bi-Hamiltonian methods [1].
The equation (1.1) is a gauge transformed form of it. We employ the above simplified equation
for convenience.
 Recently, the equation was shown to
model the propagation of nonlinear light  pulses in monomode optical fibers when certain nonlinear effects are taken into account [2].
It admits  a Lax pair which enables one to solve the initial value problem by means of the inverse scattering transform (IST) method [3].
Actually,  a few conservation laws and a bright 1-soliton solution were obtained explicitly in [3]. Subsequently, the general bright $N$-soliton solution was
constracted  with the use of the dressing method [4], where $N$ is an arbitrary positive integer. 
Note, however, that the detailed analysis of the multisoliton solutions has not been undertaken.
 For the purpose of obtaining soliton solutions in a simple manner, one can employ 
a direct approach which is now known as Hirota's direct method or bilinear transformation method [5, 6].  The method has been applied successfully to
a large number of soliton equations including the Korteweg-de Vries, NLS and sine-Gordon equations. \par
The purpose of the present paper is to construct the bright $N$-soliton of the FL DNLS equation within the framework of the bilinear formalism. In particular, we present
two different expressions of the bright $N$-soliton solution in a simple determinantal form. 
The proof of the solution 
can be obtained by a straightforward calculation where
 one relies only on a few basic formulas for determinants.
 This paper is organized as follows. In section 2, we develop a direct method of solution.  Specifically, we first transform equation (1.1) to  a
system of bilinear equations by introducing an appropriate dependent variable transformation. 
Notably, we show that one of the bilinear equations can be replaced by a {\it trilinear} equation which is easy to prove when compared with other bilinear equations.
In section 3, we present the bright $N$-soliton solution of the bilinear equations.
 It has a simple structure expressed in terms of certain determinants. 
 Subsequently,  we perform the proof of the bright $N$-soliton solution using an elementary theory of determinants 
in which Jacobi's identity  will  play a central role. Last, we provide an alternative expression of the bright $N$-soliton solution. 
The equivalence of both expressions is  discussed by referring to the properties of the Cauchy matrix. 
We emphasize that the  bright $N$-soliton solution obtained here  yields the bright $N$-soliton solution of the derivative NLS equation by replacing simply 
the time dependence of the solution. 
This connection has also been found in the context of the IST [4].
In section 4, we investigate 
the properties of the 1- and 2-soliton solutions and then analyze
the asymptotic behavior of the bright
$N$-soliton solution in the limit of infinite time.  As a result, we obtain the explicit formula for the phase shift in terms of the amplitude parameters of solitons. 
Section 5 is devoted to concluding remarks. \par
\bigskip
\leftline{\bf 2. Exact method of solution}\par
\bigskip
\noindent In this section, we develop a direct method of solution for constructing bright soliton solutions of equation (1.1)
under the boundary condition $u\rightarrow 0$ as $|x|\rightarrow \infty$. 
Specifically, we show that equation (1.1) can be transformed to  a system of bilinear equations. 
We also demonstrate that one of the bilinear equations can be replaced by a trilinear equation.
\par
\bigskip
\leftline{\it 2.1. Bilinearization}\par
\noindent The bilinearization of equation (1.1) is established by the following proposition: \par
\medskip
\noindent{\bf Proposition 2.1.}\ {\it By means of the dependent variable transformation
$$u={g\over f}, \eqno(2.1)$$
equation (1.1) can be decoupled into the following system of bilinear equations for 
the tau functions $f$ and $g$
$$D_xD_tg\cdot f=gf, \eqno(2.2)$$
$$D_tf\cdot f^*={\rm i}gg^*, \eqno(2.3)$$
$$D_xD_tf\cdot f^*={\rm i}D_xg\cdot g^*. \eqno(2.4)$$
Here, $f=f(x, t)$ and $g=g(x, t)$ are complex-valued functions of $x$ and $t$,
 and the asterisk appended to $f$ denotes complex conjugate and
the bilinear operators $D_x$ and $D_t$ are defined by
$$D_x^mD_t^nf\cdot g=\left({\partial\over\partial x}-{\partial\over\partial x^\prime}\right)^m
\left({\partial\over\partial t}-{\partial\over\partial t^\prime}\right)^n
f(x, t)g(x^\prime,t^\prime)\Big|_{ x^\prime=x,\,t^\prime=t},  \eqno(2.5)$$
where $m$ and $n$ are nonnegative integers.} \par
\bigskip
\noindent {\bf Proof.}  Substituting (2.1) into (1.1) and rewriting the resultant equation in terms of the bilinear operators,
equation (1.1) can be rewritten as
$${1\over f^2}(D_xD_tg\cdot f - gf)-{g\over f^3f^*}(f^*D_xD_tf\cdot f+2{\rm i}g^*D_xf\cdot g)=0. \eqno(2.6)$$
Inserting the identity 
$$f^*D_xD_tf\cdot f=fD_xD_tf\cdot f^*-2f_xD_tf\cdot f^*+f(D_tf\cdot f^*)_x, \eqno(2.7)$$
which can be verified by direct calculation into the second term on the left-hand side of (2.6), one modifies it in the form
$${1\over f^2}(D_xD_tg\cdot f - gf)$$
$$-{g\over f^3f^*}\Bigl\{f(D_xD_tf\cdot f^* -{\rm i}D_xg\cdot g^*)-2f_x(D_tf\cdot f^*-{\rm i}gg^*)+f(D_tf\cdot f^*-{\rm i}gg^*)_x\Bigr\}=0. \eqno(2.8)$$
By virtue of equations (2.2)-(2.4), the left-hand side of (2.8) vanishes identically. \hspace{\fill}$\Box$ \par
\bigskip
It follows from (2.1) and (2.3) that
$$|u|^2={\rm i}\,{\partial \over \partial t}\,{\rm ln}\,{f^*\over f}. \eqno(2.9)$$
The above formula gives the modulus of $u$ in terms of the tau function $f$. \par
\bigskip
\leftline{\it  2.2. Remark} \par
\noindent{\bf Proposition 2.2.} {\it The {\it trilinear} equation for $f$ and $g$,
$$f^*(g_{xt}f-g_tf_x-gf)=f_t^*(g_xf-gf_x). \eqno(2.10)$$
 is a consequence of the bilinear equations (2.2)-(2.4).} \par
\medskip
\noindent {\bf Proof.} Using the definition of the bilinear operators, one can show that
$$f^*(g_{xt}f-g_tf_x-gf)-f_t^*(g_xf-gf_x)$$
$$=f^*(D_xD_tg\cdot f-gf)-{g\over 2}\Bigl\{(D_tf\cdot f^*-{\rm i}gg^*)_x+(D_xD_tf\cdot f^*-{\rm i}D_xg\cdot g^*)\Bigr\}+g_x(D_tf\cdot f^*-{\rm i}gg^*). \eqno(2.11)$$
The right-hand side of (2.11) becomes zero by (2.2)-(2.4), which yields (2.10).
  \hspace{\fill}$\Box$ \par
\bigskip
In view of proposition 2.2, we can use equation (2.10) in place of (2.2), for example in performing the proof
of the bright  $N$-soliton solution.  We will see later in section 3.6 that the proof of (2.10)  is simplified considerably compared with that of (2.2), even if
(2.10) in not bilinear but trilinear in $f$ and $g$.  
\par
\bigskip
\leftline{\bf 3. Bright $N$-soliton solution and its proof}\par
\bigskip
\noindent In this section, we show that the tau functions $f$ and $g$ representing the bright $N$-soliton solution
admit the compact determinantal expressions.
This statement is proved by an elementary calculation using the basic formulas for determinants. Then,
we provide an alternative form of the bright $N$-soliton solution.
Last, we demonstrate that the tau functions $f$ and $g$ satisfy a trilinear equation. The implication of this 
result will be discussed briefly in conjunction with the bright $N$-soliton solution of the DNLS equation.
\par
\bigskip
\noindent{\it 3.1. Bright $N$-soliton solution} \par
\medskip
\noindent The main result in this paper is given by the following theorem. \par
\noindent{\bf Theorem 3.1.} {\it The bright $N$-soliton solution of the system of bilinear equations (2.2)-(2.4) is expressed by the
 following determinants
$$f=|D|, \quad g=\begin{vmatrix} D & {\bf z}_t^T\\ {\bf 1} & 0\end{vmatrix}. \eqno(3.1)$$
Here, $D$ is an $N\times N$ matrix and ${\bf z}, {\bf z}_t$ and ${\bf 1}$ are $N$-component row vectors defined below and 
the symbol $T$  denotes the transpose:
$$D=(d_{jk})_{1\leq j,k\leq N}, \quad d_{jk}={z_jz_k^*-{\rm i}p_k^*\over p_j+p_k^*}, \quad z_j={\rm exp}\left(p_jx+{1\over p_j}t+\zeta_{j0}\right), \eqno(3.2a)$$
$${\bf z}=(z_1, z_2, ..., z_N), \quad {\bf z}_t=(z_1/p_1, z_2/p_2, ..., z_N/p_N), \quad {\bf 1}=(1, 1, ..., 1), \eqno(3.2b)$$ 
where  $p_j$ and $\zeta_{j0}\ (j=1, 2, ..., N)$ are arbitrary complex parameters. } \par
\bigskip
\noindent The bright $N$-soliton solution is parameterized $2N$ complex parameters $p_j$ and $\zeta_{j0}\ (j=1, 2, ..., N)$. 
The  parameters $p_j$ determine the amplitude and velocity of the solitons whereas the parameters $\zeta_{j0}$ determine   the 
 phase of the solitons.  The conditions  
$p_j+p_k^*\not=0$ for all $j$ and $k$ and $p_j\not=p_k$ for $j\not=k$  may be imposed on the parameters 
to assure the regularity of the solution. 
We point out that the $N$-soliton formula given by Theorem 3.1 
 can be shown to take the same form as that obtained in [4] by changing the
soliton parameters $p_j$ and $\zeta_{j0}\ (j=1, 2, ..., N)$.
\par
\bigskip
\noindent{\it 3.2. Notation and basic formulas for determinants} \par
\medskip
\noindent First, we define the definition of matrices associated with the bright $N$-soliton solution and then provide some basic 
formulas for determinants. The following bordered matrices appear frequently in our analysis:
$$D({\bf a}; {\bf b})=\begin{pmatrix} D & {\bf b}^T\\ {\bf a} & 0\end{pmatrix},\quad
D({\bf a},{\bf b};{\bf c},{\bf d})=\begin{pmatrix} D &{\bf c}^T & {\bf d}^T \\ {\bf a} & 0 &0\\
                                                                            {\bf b} & 0& 0 \end{pmatrix}, \eqno(3.3)$$
where ${\bf a}, {\bf b}, {\bf c}$ and {\bf d} are $N$ component row vectors. 
 Let  $D_{jk}$ be the cofactor of the element $d_{jk}$. The following formulas are well known in the theory of determinants:
$${\partial\over\partial x}|D|=\sum_{j,k=1}^N{\partial d_{jk}\over\partial x}D_{jk}, \eqno(3.4)$$
$$\begin{vmatrix} D & {\bf a}^T\\ {\bf b} & z\end{vmatrix}=|D|z-\sum_{j,k=1}^ND_{jk}a_jb_k,  \eqno(3.5)$$
$$|D({\bf a}, {\bf b}; {\bf c}, {\bf d})||D|= |D({\bf a}; {\bf c})||D({\bf b}; {\bf d})|-|D({\bf a}; {\bf d})||D({\bf b}; {\bf c})|. \eqno(3.6)$$
Formula (3.4) is the differentiation rule of the determinant and (3.5) is the expansion formula for a bordered determinant
with respect to the last row and last column.
Formula (3.6) is Jacobi's identity.   \par
\bigskip
\noindent{\it 3.3. Differentiation rules and related formulas} \par
\noindent In terms of the notation (3.3), the tau functions $f$ and $g$ can be written as $f=|D|$ and $g=|D({\bf 1};{\bf z}_t)|$, respectively.
The differentiation rules of the tau functions with respect to $t$ and $x$ are given by the following formulas: \par
\medskip
\noindent{\bf Lemma 3.1.} \par
$$f_t=-|D({\bf z}_t^*;{\bf z}_t)|, \eqno(3.7)$$
$$f_x=-|D({\bf z}^*;{\bf z})|, \eqno(3.8)$$
$$f_{xt}=-|D({\bf z}_t^*;{\bf z})|-|D({\bf z}^*;{\bf z}_t)|+|D({\bf z}^*,{\bf z}_t^*;{\bf z},{\bf z}_t)|, \eqno(3.9)$$
$$g_t=|D({\bf 1};{\bf z}_{tt})|, \eqno(3.10)$$
$$g_x=|D({\bf 1};{\bf z})|-|D({\bf 1},{\bf z}^*;{\bf z}_t,{\bf z})|, \eqno(3.11)$$
$$g_{xt}=|D({\bf 1};{\bf z}_t)|-|D({\bf 1},{\bf z}^*;{\bf z}_{tt},{\bf z})|. \eqno(3.12)$$
\medskip
\noindent {\bf Proof.}  We prove (3.7). Applying formula (3.4) to $f$ given by (3.1) with (3.2a), one obtains
\begin{align}
f_t &=\sum_{j,k=1}^ND_{jk}{z_jz_k^*\over p_jp_k^*} \notag \\
    &=\sum_{j,k=1}^ND_{jk}z_{j,t}z_{k,t}^*, \notag
\end{align}
where in passing to the second line, use has been made of the relations $z_j/p_j=z_{j,t},\ z_k^* p_k^*=z_{k,t}^*$. Referring to  formula (3.5) with $z=0$ and 
taking into account the notation (3.3), the above expression
becomes the right-hand side of (3.7). Formulas (3.8)-(3.12) can be proved in the same way if one uses (3.4), (3.5) and the relation ${\bf z}_{xt}={\bf z}$ as well
as some basic properties of determinants. \hspace{\fill}$\Box$ \par
\medskip
\noindent The complex conjugate expressions of the tau functions $f$ and $g$ and their derivatives are expressed as follows: \par
\medskip
\noindent{\bf Lemma 3.2.}\par
$$f^*=|\bar D|, \qquad \bar D={D^*}^T, \eqno(3.13) $$
$$f_t^*=-|\bar D({\bf z}_t^*;{\bf z}_t)|, \eqno(3.14)$$
$$f_x^*=-|\bar D({\bf z}^*;{\bf z})|, \eqno(3.15)$$
$$f_{xt}^*=-|\bar D({\bf z}^*;{\bf z}_t)|-|\bar D({\bf z}_t^*;{\bf z})|+|\bar D({\bf z}^*,{\bf z}_t^*;{\bf z},{\bf z}_t)|, \eqno(3.16)$$
$$g^*=|\bar D({\bf z}_t^*;{\bf 1})|, \eqno(3.17)$$
$$g_x^*=|\bar D({\bf z}^*;{\bf 1})|-|\bar D({\bf z}_t^*,{\bf z}^*;{\bf 1},{\bf z})|, \eqno(3.18)$$
{\it where $\bar D=(\bar d_{jk})_{1\leq j,k\leq N}$ is an $N\times N$ matrix with elements $\bar d_{jk}=d_{jk}+{\rm i}$.\it} \par
\medskip
\noindent {\bf Proof.} It follows from (3.2a) that $d_{jk}^*=d_{kj}+{\rm i}$ or equivalently,  $D^*=\bar D^T$, which proves
(3.13) since $f^*=|D^*|=|\bar D^T|=|\bar D|$. Formulas (3.14)-(3.18) can be proved in the same way. \hspace{\fill}$\Box$ \par
\noindent The formulas  below are used to reduce the proof of the bright $N$-soliton solution to Jacobi's identity. \par
\noindent{\bf Lemma 3.3.} \par
$$|D({\bf 1};{\bf z})|=|\bar D({\bf 1};{\bf z})|, \eqno(3.19)$$
$$|D({\bf 1};{\bf z}_t)|=|\bar D({\bf 1};{\bf z}_t)|=(-1)^{N-1}c|\bar D(\tilde {\bf p}^*;{\bf z})|, \eqno(3.20)$$
$$|D({\bf z}^*;{\bf z})|=(-1)^{N-1}c|\bar D({\bf z}_t^*;{\bf z}_x)|, \eqno(3.21)$$
$$|D({\bf z}^*;{\bf z}_t)|=(-1)^{N-1}c|\bar D({\bf z}_t^*;{\bf z})|, \eqno(3.22)$$
$$|D({\bf z}^*;{\bf z}_{tt})|=(-1)^{N-1}c|\bar D({\bf z}_t^*;{\bf z}_t)|, \eqno(3.23)$$
$$|D({\bf z}^*,{\bf 1};{\bf z},\tilde{\bf p})|=(-1)^Nc|\bar D({\bf z}_t^*,\tilde{\bf p}^*;{\bf z}_x,{\bf 1})|, \eqno(3.24)$$
$$|D({\bf 1},{\bf z}_t^*;{\bf z},{\bf z}_t)|=|\bar D({\bf 1},{\bf z}_t^*;{\bf z},{\bf z}_t)|, \eqno(3.25)$$
{\it where $\tilde{\bf p}=(1/p_1, 1/p_2, ..., 1/p_N)$ and ${\bf z}_x=(p_1z_1, p_2z_2, ..., p_Nz_N)$ are $N$ component row vectors and $c=\prod_{j=1}^N(p_j^*/p_j)$. \it} \par
\medskip
\noindent {\bf Proof.} We prove (3.21) only. The other formulas can be proved following the same procedure as that described below.
 First, using the key relation which follows from (3.2a) and the definition of $\bar d_{jk}$, 
$$d_{jk}-{z_jz_k^*\over p_j}=-{p_k^*\over p_j}(d_{jk}+{\rm i})=-{p_k^*\over p_j}\bar d_{jk}, \eqno(3.26)$$
the left-hand side of (3.21) can be modified into the form
$$ |D({\bf z}^*;{\bf z})|=\begin{vmatrix} \left(-{p_k^*\over p_j}\bar d_{jk}\right) & {\bf z}^T\\ {\bf z}^* & 0\end{vmatrix}, \eqno(3.27)$$
after multiplying the $(N+1)$th row by $-z_j/p_j$ and  then adding the resultant expression to the $j$th row for $j= 1, 2, ..., N$. Formula (3.21)
follows from (3.27) if one extracts the factors $1/p_j$ and $-p_k^*$ from the $j$th row and $k$th column, respectively for $j, k = 1, 2, ..., N$ and the factor
$-1$ from the $(N+1)$th row. \hspace{\fill}$\Box$ \par
\bigskip
\noindent{\it 3.4. Proof of the bright $N$-soliton solution}\par
\medskip
\noindent{\it 3.4.1. Proof of (2.2)}. Let  
$$P_1=D_xD_tg\cdot f-gf=g_{xt}f-g_xf_t-g_tf_x+gf_{xt}-gf. \eqno(3.28) $$
Substituting formulas (3.7)-(3.12) into this expression  and applying Jacobi's identity to a term $|D({\bf 1},{\bf z}^*;{\bf z}_{tt},{\bf z})||D|$,
$P_1$ reduces to
$$P_1=P_{11}+P_{12}+P_{13}, \eqno(3.29a)$$
with
$$P_{11}=|D({\bf 1};{\bf z})||D({\bf z}_t^*;{\bf z}_t)|-|D({\bf 1};{\bf z}_t)||D({\bf z}_t^*;{\bf z})|, \eqno(3.29b)$$
$$P_{12}=|D({\bf 1};{\bf z}_t)||D({\bf z}^*,{\bf z}_t^*;{\bf z},{\bf z}_t)| -|D({\bf z}_t^*;{\bf z}_t)||D({\bf 1},{\bf z}^*;{\bf z}_t,{\bf z})|, \eqno(3.29c)$$
$$P_{13}=|D({\bf 1};{\bf z})||D({\bf z}^*;{\bf z}_{tt})|-|D({\bf 1};{\bf z}_t)||D({\bf z}^*;{\bf z}_t)|. \eqno(3.29d)$$
Referring  to Jacobi's identity, $P_{11}$ becomes
$$P_{11}=|D||D({\bf 1},{\bf z}_t^*;{\bf z},{\bf z}_t)|. \eqno(3.30)$$
Consider the identity 
$$\begin{vmatrix} |D({\bf 1};{\bf z}_t)| & |D({\bf 1};{\bf z}_t)| & |D({\bf 1};{\bf z})| \\
|D({\bf z}_t^*;{\bf z}_t)| & |D({\bf z}_t^*;{\bf z}_t)| & |D({\bf z}_t^*;{\bf z})| \\
|D({\bf z}^*;{\bf z}_t)| & |D({\bf z}^*;{\bf z}_t)| & |D({\bf z}^*;{\bf z})|   \end{vmatrix}=0, \eqno(3.31)$$
which follows immediately since the first two columns of the determinant coincide. Expanding the above determinant with respect to the
first column and using Jacobi's identity, one finds that
$$P_{12}=|D({\bf z}^*;{\bf z}_t)||D({\bf 1},{\bf z}_t^*;{\bf z},{\bf z}_t)|. \eqno(3.32)$$
Gathering up three terms (3.29d), (3.30) and (3.32), $P_1$ simplifies to
$$P_1=\bigl\{|D|+|D({\bf z}^*;{\bf z}_t)|\bigr\}|D({\bf 1},{\bf z}_t^*;{\bf z},{\bf z}_t)|+|D({\bf 1};{\bf z})||D({\bf z}^*;{\bf z}_{tt})|-|D({\bf 1};{\bf z}_t)||D({\bf z}^*;{\bf z}_t)|. \eqno(3.33)$$
It follows from formula (3.5) that
$$|D|+|D({\bf z}^*;{\bf z}_t)|=\begin{vmatrix} D & {\bf z}_t^T\\ {\bf z}^* & 1\end{vmatrix}=(-1)^Nc|\bar D|, \eqno(3.34)$$
where in passing to the last line, the procedure used for deriving lemma 3.3 has been applied.
 If one substitutes this result together with (3.19), (3.20), (3.23) and (3.25) into $P_1$, one can
recast $P_1$ into the form
$$P_1=(-1)^{N-1}c\bigl\{-|\bar D||\bar D({\bf 1},{\bf z}_t^*;{\bf z},{\bf z}_t)|+|\bar D({\bf 1};{\bf z})||\bar D({\bf z}_t^*;{\bf z}_t)|-|\bar D({\bf 1};{\bf z}_t)||\bar D({\bf z}_t^*;{\bf z})|\bigr\}, \eqno(3.35)$$
which becomes zero by virtue of Jacobi's identity. \hspace{\fill}$\Box$ \par
\medskip
\noindent{\it 3.4.2. Proof of (2.3)}. Let 
 $$P_2=D_tf\cdot f^*-{\rm i}gg^*=f_tf^*-ff_t^*-{\rm i}gg^*. \eqno(3.36)$$
Substituting formulas (3.7), (3.13), (3.14) and (3.17) into this expression, $P_2$ becomes
$$P_2=-|D({\bf z}_t^*;{\bf z}_t)||\bar D|+|D||\bar D({\bf z}_t^*;{\bf z}_t)|-{\rm i}|\bar D({\bf z}_t^*;{\bf 1})||D({\bf 1};{\bf z}_t)|. \eqno(3.37)$$
The formulas below can be derived by using the definition of the matrix $\bar D$ from (3.13):
$$|\bar D|=|D|-{\rm i}|D({\bf 1};{\bf 1})|, \eqno(3.38)$$
$$|\bar D({\bf z}_t^*;{\bf 1})|=|D({\bf z}_t^*;{\bf 1})|, \eqno(3.39)$$
$$|\bar D({\bf z}_t^*;{\bf z}_t)|=|D({\bf z}_t^*;{\bf z}_t)|-{\rm i}|D({\bf z}_t^*,{\bf 1};{\bf z}_t,{\bf 1})|. \eqno(3.40)$$
After introducing (3.38)-(3.40) into (3.37), $P_2$ reduces to
$$P_2={\rm i}\Bigl\{|D({\bf z}_t^*;{\bf z}_t)||D({\bf 1};{\bf 1})|-|D||D({\bf z}_t^*,{\bf 1};{\bf z}_t,{\bf 1})|-|D({\bf z}_t^*;{\bf 1})||D({\bf 1};{\bf z}_t)|\Bigl\}. \eqno(3.41)$$
One sees that $P_2$ becomes zero by Jacobi's identity. \par
\medskip
\noindent{\it 3.4.3. Proof of (2.4)}.\par
\medskip
\noindent Instead of proving (2.4) directly, we differentiate (2.3) by $x$ and add the resultant expression to (2.4)
and then prove the equation $P_3=0$, where
$$P_3=f_{xt}f^*-f_xf_t^*-{\rm i}g_xg^*. \eqno(3.42) $$
Substituting (3.8), (3.9), (3.11), (3.13), (3.14) and (3.17) into (3.42), $P_3$ becomes
$$P_3=\bigl\{-|D({\bf z}_t^*;{\bf z})|-|D({\bf z}^*;{\bf z}_t)|+|D({\bf z}^*,{\bf z}_t^*;{\bf z},{\bf z}_t)|\bigr)\}|\bar D|
-|D({\bf z}^*;{\bf z})||\bar D({\bf z}_t^*;{\bf z}_t)|$$
$$-{\rm i}\bigl\{|D({\bf 1};{\bf z})|-|D({\bf 1},{\bf z}^*;{\bf z}_t,{\bf z})|\bigr\}|\bar D({\bf z}_t^*;{\bf 1})|. \eqno(3.43)$$
Using the cofactor expansion of the determinant $|D|$, 
$$\sum_{k=1}^Nd_{jk}D_{jk}=|D|,\qquad \sum_{j=1}^Nd_{jk}D_{jk}=|D|, \eqno(3.44)$$
 one obtains 
$$\sum_{j,k=1}^N\left({1\over p_j}+{1\over p_k^*}\right)d_{jk}D_{jk}=\sum_{j=1}^N\left({1\over p_j}+{1\over p_j^*}\right)|D|. \eqno(3.45)$$
The definition of $d_{jk}$ from (3.2a) gives $\left({1\over p_j}+{1\over p_k^*}\right)d_{jk}=z_{j,t}z_{k,t}^*-{\rm i}\tilde p_j$, which, after substituting  into (3.45) 
 and using formula (3.5), yields the relation
$$-|D({\bf z}_t^*;{\bf z}_t)|+{\rm i}|D({\bf 1};\tilde {\bf p})|=\sum_{j=1}^N\left({1\over p_j}+{1\over p_j^*}\right)|D|. \eqno(3.46)$$
Invoking the relation $D^*=\bar D^T$, the complex conjugate of this expression can be written in the form
$$-|\bar D({\bf z}_t^*;{\bf z}_t)|-{\rm i}|\bar D(\tilde {\bf p}^*;{\bf 1})|=\sum_{j=1}^N\left({1\over p_j}+{1\over p_j^*}\right)|\bar D|. \eqno(3.47)$$
The following formula can be derived if one differentiates (3.46) by $x$ and notes the relation ${\bf z}_{xt}={\bf z}$
$$-|D({\bf z}_t^*;{\bf z})|-|D({\bf z}^*;{\bf z}_t)|+|D({\bf z}^*,{\bf z}_t^*;{\bf z},{\bf z}_t)|-{\rm i}|D({\bf z}^*,{\bf 1};{\bf z},\tilde {\bf p})|$$
$$=-\sum_{j=1}^N\left({1\over p_j}+{1\over p_j^*}\right)|D({\bf z}^*;{\bf z})|. \eqno(3.48)$$
Multiply (3.48) by $|\bar D|$ and use (3.47) to obtain
$$\bigl\{-|D({\bf z}_t^*;{\bf z})|-|D({\bf z}^*;{\bf z}_t)|+|D({\bf z}^*,{\bf z}_t^*;{\bf z},{\bf z}_t)|\bigr\}|\bar D|$$
$$={\rm i}|D({\bf z}^*,{\bf 1};{\bf z},\tilde {\bf p})||\bar D|+\bigl\{|\bar D({\bf z}_t^*;{\bf z}_t)|+{\rm i}|\bar D(\tilde {\bf p}^*;{\bf 1})|\bigr\}|D({\bf z}^*;{\bf z})|, \eqno(3.49)$$
which, substituted into $P_3$, gives
$$P_3={\rm i}\Bigl[|D({\bf z}^*,{\bf 1};{\bf z},\tilde {\bf p})||\bar D|+|\bar D(\tilde {\bf p}^*;{\bf 1})||D({\bf z}^*;{\bf z})|
-\bigl\{|D({\bf 1};{\bf z})|-|D({\bf 1},{\bf z}^*;{\bf z}_t,{\bf z})|\bigr\}|\bar D({\bf z}_t^*;{\bf 1})|\Bigr]. \eqno(3.50)$$
Differentiation of (3.20) with respect to $x$ yields the relation
$$|D({\bf 1};{\bf z})|-|D({\bf 1},{\bf z}^*;{\bf z}_t,{\bf z})|=(-1)^{N-1}c|\bar D(\tilde {\bf p}^*;{\bf z}_x)|. \eqno(3.51)$$
If one substitutes (3.21), (3.24) and (3.51) into (3.50), $P_3$ simplifies to
$$P_3={\rm i}(-1)^Nc\Bigl\{|\bar D({\bf z}_t^*,\tilde{\bf p}^*;{\bf z}_x,{\bf 1})||\bar D|-|\bar D(\tilde {\bf p}^*;{\bf 1})||\bar D({\bf z}_t^*;{\bf z}_x)|
+|\bar D(\tilde {\bf p}^*;{\bf z}_x)||\bar D({\bf z}_t^*;{\bf 1})|\Bigr\}. \eqno(3.52)$$
It turns out that $P_3$ becomes zero by Jacobi's identity. \hspace{\fill}$\Box$ \par
\medskip
\noindent{\it 3.5. An alternative expression of the bright $N$-soliton solution}\par
\medskip
\noindent Here, we present an alternative expression of the bright $N$-soliton solution of equation (1.1). It is 
expressed by the following theorem:\par
\medskip
\noindent{\bf Theorem 3.2.} {\it  The tau functions $f^\prime$ and $g^\prime$ given below satisfy the system of bilinear equations (2.2)-(2.4):
$$f^\prime=\begin{vmatrix}A & I\\
                          -I & B \end{vmatrix}, \qquad g^\prime=\begin{vmatrix}A & I & {\bf y}_t^T \\
                          -I & B & {\bf 0}^T\\
                         {\bf 0} & {\bf 1} & 0 \end{vmatrix}. \eqno(3.53)$$
Here, $A, B$ and $I$ are $N\times N$ matrices and ${\bf y}$ and ${\bf y}_t$ are $N$ component row vectors defined respectively by
$$A=(a_{jk})_{1\leq j,k\leq N}, \qquad a_{jk}={y_jy_k^*\over q_j+q_k^*},\qquad y_j={\rm exp}\left(q_jx+{1\over q_j}\,t+\eta_{j0}\right), \eqno(3.54a)$$
$$B=(b_{jk})_{1\leq j,k\leq N}, \qquad b_{jk}={{\rm i}q_k\over q_j^*+q_k}, \eqno(3.54b)$$
$$I=(\delta_{jk})_{1\leq j,k\leq N}: N\times N\ {\it unit\ matrix}, \eqno(3.54c)$$
$${\bf y}=(y_1, y_2, ..., y_N), \qquad {\bf y}_t=(y_1/q_1, y_2/q_2, ..., y_N/q_N). \eqno(3.54d)$$
where  $q_j$ and $\eta_{j0}\ (j=1, 2, ..., N)$ are arbitrary complex parameters.} \par
\bigskip
\noindent The proof of theorem 3.2 parallels theorem 3.1 and hence omitted. Instead, we provide an alternative proof.
To this end, we first  establish the following proposition: \par
\medskip
\noindent{\bf Proposition 3.1.} {\it Under the parameterization $q_j=-p_j^*, \eta_{j0}=-\zeta_{j0}^*+{\rm ln}\,c_j\ (j=1, 2, ..., N)$, the 
tau functions $f, g, f^\prime$ and $g^\prime$ satisfy the relations
$$f^\prime=(-1)^N|A|f, \qquad g^\prime=-c^\prime|A|g, \eqno(3.55)$$
where
$$c^\prime=\prod_{j=1}^N{q_j\over q_j^*}, \qquad c_j={\prod_{l=1}^N(q_j+q_l^*)\over \prod_{\substack{l=1\\(l\not=j)}}^N(q_j-q_l)}, \qquad j= 1, 2, ..., N, \eqno(3.56)$$ 
and the conditions $q_j+q_k^*\not=0$ for all $j$ and $k$ and $q_j\not=q_k$ for $j\not=k$ are imposed on the parameters.}\par
\medskip
\noindent {\bf Proof.} By means of the operation of matrix multiplication, one can show that
$$f^\prime=|I+AB|=|A||A^{-1}+B|, \qquad g^\prime=\begin{vmatrix} I+AB & {\bf y}_t^T\\
                                                                  {\bf 1} & 0\end{vmatrix}
                                                                  =|A|\begin{vmatrix} A^{-1}+B & A^{-1}{\bf y}_t^T\\
                                                                  {\bf 1} & 0\end{vmatrix}. \eqno(3.57)$$
The inverse of the Cauchy matrix $A$ exists due to the conditions imposed on the parameters. It reads [7]
$$A^{-1}=\left({c_j^*c_k\over q_j^*+q_k}{1\over y_j^*y_k}\right)_{1\leq j,k\leq N}. \eqno(3.58)$$
The specified parametrization also leads to the relation $y_j=c_j/z_j^*$, which, substituted into $A^{-1}$ and $B$, gives
$$A^{-1}=\left(-{z_jz_k^*\over p_j+p_k^*}\right)_{1\leq j,k\leq N},\qquad B=\left({{\rm i}p_k^*\over p_j+p_k^*}\right)_{1\leq j,k\leq N}. \eqno(3.59)$$
It follows from  (3.2a) and (3.59) that $A^{-1}+B=-D$.  If one introduces this relation into $f^\prime$ from (3.57), one finds the first relation of (3.55). 
To proceed, note that the $j$th element of the column vector $A^{-1}{\bf y}_t^T$ is given by
$$\left(A^{-1}{\bf y}_t^T\right)_j=\sum_{m=1}^N(A^{-1})_{jm}y_{m,t}={c_j^*\over y_j^*}\sum_{m=1}^N{1\over q_m}{\prod_{\substack{l=1\\(l\not=j)}}^N(q_m+q_l^*)\over \prod_{\substack{l=1\\(l\not=m)}}^N(q_m-q_l)}.\eqno(3.60)$$
The sum with respect to $m$ turns out to be $(-1)^{N-1}\prod_{l=1}^N(q_l/q_l^*)/q_j^*$ by using Euler's formula. Hence, (3.60) reduces to $(-1)^Nc^\prime z_{j,t}$ upon rewriting the resultant expression in terms of $z_j$. Thus,
$g^\prime$ from (3.57) becomes
$$g^\prime=(-1)^Nc^\prime|A|\begin{vmatrix} -D & {\bf z}_t^T\\
                             {\bf 1} & 0\end{vmatrix}=-c^\prime|A|g,$$
giving rise to the second relation of (3.55). \hspace{\fill}$\Box$ \par
\medskip
\noindent{\bf Proposition 3.2.} {\it The tau functions $f^\prime$ and $g^\prime$ from (3.55)   satisfy the system of bilinear equations (2.2)-(2.4) if $f$ and $g$ satisfy the same
system of equations.}\par
\medskip
\noindent {\bf Proof.}\ Let $P_1^\prime, P_2^\prime$ and $P_3^\prime$ be
$$P_1^\prime=D_xD_tg^\prime\cdot f^\prime-g^\prime f^\prime={g_{xt}}^\prime f^\prime-g^\prime_xf^\prime_t-g^\prime_tf^\prime_x+g^\prime f^\prime_{xt}-g^\prime f^\prime, \eqno(3.61) $$
$$P^\prime_2=D_tf^\prime\cdot {f^\prime}^*-{\rm i}g^\prime {g^\prime}^*=f^\prime_t{f^\prime}^*-f^\prime{f^\prime}_t^*-{\rm i}g^\prime {g^\prime}^*, \eqno(3.62)$$
$$P^\prime_3=f^\prime_{xt}{f^\prime}^*-f^\prime_x{f^\prime}_t^*-{\rm i}g^\prime_x{g^\prime}^*, \eqno(3.63) $$
respectively. First, note that the determinant $|A|$ from (3.54a) can be modified in the form
\begin{align}
|A|&=\prod_{j=1}^Ny_j\prod_{k=1}^Ny_k^*\left|\left({1\over q_j+q_k^*}\right)\right| \notag \\
&={\rm exp}\left[\sum_{j=1}^N\left\{(q_j+q_j^*)x+\left({1\over q_j}+{1\over q_j^*}\right)t+\eta_{j0}+\eta_{j0}^*\right\}\right]\left|\left({1\over q_j+q_k^*}\right)\right|.\tag{3.64} 
\end{align}
It immediately follows from this expression that
$$D_xD_t|A|\cdot |A|=0. \eqno(3.65)$$
Substitute (3.55) into (3.61)-(3.63) and use the relation $|c^\prime|=1$ to obtain 
$$P_1^\prime=(-1)^{N-1}c^\prime\bigl\{|A|^2(D_xD_tg\cdot f-gf)+(D_xD_t|A|\cdot |A|)gf\bigr\}, \eqno(3.66)$$
$$P_2^\prime=|A|^2(D_tf\cdot f^*-{\rm i}gg^*), \eqno(3.67)$$
$$P_3^\prime=|A||A|_x(D_tf\cdot f^*-{\rm i}gg^*)+|A|^2(f_{xt}f^*-f_xf_t^*-{\rm i}g_xg^*)+{1\over 2}(D_xD_t|A|\cdot |A|)ff^*. \eqno(3.68)$$
Above three expressions vanish identically by virtue of (2.3), (2.4) $P_3=0$ with $P_3$ given by (3.42),  and (3.65). \hspace{\fill}$\Box$ \par
\bigskip
Propositions 3.1 and 3.2 lead to an alternative expression of the bright $N$-soliton solution of equation (1.1) in terms of the tau functions $f^\prime$ and $g^\prime$:
$$u=(-1)^{N-1}c^\prime\,{g^\prime\over f^\prime},\qquad |u|^2={\rm i}\, {\partial\over \partial t}\,{\rm ln}\,{{f^\prime}^*\over f^\prime}. \eqno(3.69)$$
Note that if $u$ satisfies equation (1.1) then $cu$ with $c$ being a complex constant satisfies the equation as well if $|c|=1$. Hence, the factor
$(-1)^{N-1}c^\prime$ in the above expression for $u$  is irrelevant and it  may be replaced  simply  by 1. \par
\medskip
\noindent{\it 3.6. Remark}\par
\medskip
\noindent As already mentioned in section 2.2, one of the bilinear equations (2.2)-(2.4) can be replaced by the trilinear equation (2.10).
Here, we show that the tau functions (3.1) for the bright $N$-soliton solution 
satisfy equation (2.10).  The proof is quite  simple. Indeed, referring to (3.7), (3.8), (3.10)-(3.14) and (3.23) and
using Jacobi's identity, we can derive the following relations
$$g_{xt}f-g_tf_x-gf=(-1)^{N-1}c|D({\bf 1};{\bf z})||\bar D({\bf z}_t^*;{\bf z}_t)|=(-1)^Nc|D({\bf 1};{\bf z})|f_t^*, \eqno(3.70)$$
$$g_xf-gf_x=(-1)^Nc|D({\bf 1};{\bf z})||\bar D|=(-1)^Nc|D({\bf 1};{\bf z})|f^*. \eqno(3.71)$$
Upon substituting (3.70) and (3.71) into (2.10), one can confirm that equation (2.10 ) holds identically. 
Note that  expressions (3.70) and (3.71) have a common factor $|D({\bf 1};{\bf z})|$. 
\par
Let $q=(g/f)_x$. It then follows from (3.71) that
$$q=(-1)^Nc\,{|D({\bf 1};{\bf z})|f^*\over f^2}. \eqno(3.72)$$
The form of $q$ coincides perfectly with the $N$-soliton solution of the DNLS equation
$${\rm i}q_t+q_{xx}+2{\rm i}(|q|^2q)_x=0. \eqno(3.73)$$
Actually, if we replace the time dependence of $z_j$ in (3.2a) as ${\rm i}p_j^2t$ instead of $(1/p_j)t$, then
the $N$-soliton solution (3.72) with the tau function $f$ from (3.1) satisfies the DNLS equation. See, for example [8-10].
For completeness, we reproduce the 1-soliton  solution $q_1$ as well as the square of its modulus $|q_1|^2$.
To this end, we put $p_1=a_1+{\rm i}b_1$ and $\zeta_{10}=\theta_{10}+{\rm i}\chi_{10}$.  Then, the expression (3.72) with $N=1$ yields
$$q_1= {2a_1(a_1-{\rm i}b_1)\over a_1+{\rm i}b_1}\,{{\rm e}^{\theta_1+{\rm i}\chi_1}({\rm e}^{2\theta_1}-b_1+{\rm i}a_1)\over ({\rm e}^{2\theta_1}-b_1-{\rm i}a_1)^2}, \eqno(3.74a)$$
$$ \theta_1=a_1(x-2b_1t)+\theta_{10}, \qquad \chi_1=b_1x+(a_1^2-b_1^2)t+\chi_{10}, \eqno(3.74b)$$
$$|q_1|^2={2a_1^2\over \sqrt{a_1^2+b_1^2}}\,{1\over \cosh\,2(\theta_1+\delta_1)-{b_1\over \sqrt{a_1^2+b_1^2}}},
\qquad {\rm e}^{2\delta_1}={1\over \sqrt{a_1^2+b_1^2}},\eqno(3.75)$$
Let $A_1$ and $c_1$ be the amplitude and velocity of $|q_1|$, respectively. We then find from (3.75) that
$$A_1=\sqrt{\sqrt{c_1^2+4a_1^2}+c_1}, \qquad c_1=2b_1. \eqno(3.76)$$
Note that if $b_1>0$, the soliton propagates to the right whereas if $b_1<0$, it propagates to the left. 
This fact implies that both the overtaking and head-on collisions are possible in the interaction process of two solitons.
It is interesting that in the  case of $b_1>0 (b_1<0)$, the amplitude of the soliton is an increasing (a decreasing) function of the velocity.
 \par
 The result (3.72) also follows from the view point of the IST. Indeed, if we identify $q$ with $u_x$, then the $x$-part of the Lax pair for the FL DNLS equation
 coincides with that of the DNLS equation [2]. It turns out that the bright $N$-soliton solution of the DNLS equation can be derived from that of the
 FL DNLS equation via the relation $q=u_x$ and replacing the time dependence of the exponential functions $z_j$, as demonstrated here. \par 
\bigskip
\newpage
\noindent{\bf 4. Properties of the bright $N$-soliton solution}\par
\bigskip
\noindent In this section, we investigate the properties of the bright $N$-soliton solution particularly focusing on the 1- and 2-soliton solutions
and then address the asymptotic behavior of the $N$-soliton solution for large time. To this end,
 we first parametrize the complex constants $p_j$ and $\zeta_{j0}$ by the real constants $a_j, b_j, \theta_{j0}$ and $\chi_{j0}$ as
$$p_j=a_j+{\rm i}b_j, \qquad \zeta_{j0}=\theta_{j0}+{\rm i}\chi_{j0}, \qquad a_j>0, \qquad j=1, 2, ..., N,\eqno(4.1)$$
and introduce the new independent variables $\theta_j$ and $\chi_j$ in accordance with the relations
$$\quad \theta_j=a_j(x+c_jt)+\theta_{j0}, \qquad c_j={1\over a_j^2+b_j^2}, \qquad j=1, 2, ..., N, \eqno(4.2a)$$
$$\chi_j=b_j(x-c_jt)+\chi_{j0}, \qquad j=1, 2, ..., N. \eqno(4.2b)$$
In terms of these variables, $z_j$ is expressed as
$$z_j={\rm e}^{\theta_j+{\rm i}\chi_j}, \qquad j=1, 2, ..., N. \eqno(4.2c)$$
\medskip
\noindent{\it 4.1. Bright 1-soliton solution}\par
\medskip
\noindent The tau functions $f=f_1$ and $g=g_1$ corresponding to the bright 1-soliton solution follows from (3.1) and (3.2) with $N=1$. They read
$$f_1={z_1z_1^*-{\rm i}p_1^*\over p_1+p_1^*},\qquad g_1=-{z_1\over p_1}. \eqno(4.3)$$
In terms of the new parameters defined by (4.1) and (4.2), the 1-soliton solution $u_1=u_1(\theta_1,\chi_1)$ takes the form of an envelope soliton
$$u_1=-{p_1+p_1^*\over p_1}\,{z_1\over z_1z_1^*-{\rm i}p_1^*}=-{2a_1\over a_1+{\rm i}b_1}\,{{\rm e}^{\theta_1+{\rm i}\chi_1}\over {\rm e}^{2\theta_1}-b_1-{\rm i}a_1}.\eqno(4.4)$$
The square of the modulus of  $u_1$ can be written as
$$|u_1|^2={2a_1^2\over (a_1^2+b_1^2)^{3\over 2}}\,{1\over \cosh\,2(\theta_1+\delta_1)-{b_1\over \sqrt{a_1^2+b_1^2}}}, 
\qquad {\rm e}^{2\delta_1}={1\over \sqrt{a_1^2+b_1^2}},\eqno(4.5a)$$
showing that $|u_1|$ has the amplitude $A_1$  given  by
$$A_1={\sqrt{2\,(\sqrt{a_1^2+b_1^2}+b_1)}\over \sqrt{ a_1^2+b_1^2}}. \eqno(4.5b)$$
It represents a localized pulse (or a bright soliton) moving to the left at a constant velocity $c_1=1/(a_1^2+b_1^2)$. 
The amplitude-velocity relation follows immediately from (4.5b) to give 
$$A_1=\sqrt{2(\sqrt{c_1}+b_1c_1)}. \eqno(4.6)$$ 
See Figure 1. 
The characteristic of the solution depends crucially on the signature of the parameter $b_1$, as we shall see now.
\par
\medskip
\begin{center}
\includegraphics[width=10cm]{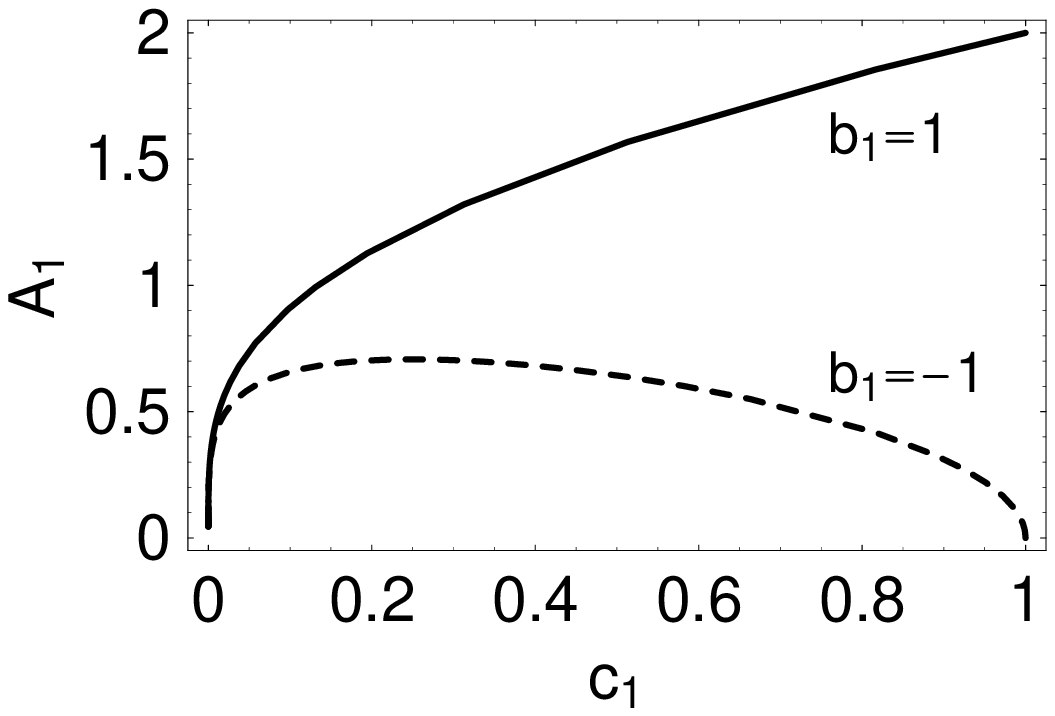}
\end{center}
{\bf Figure 1.} Amplitude-velocity relation for $b_1=1$ (solid line) and $b_1=-1$ (broken line). \par
\medskip
For a fixed value of $b_1$, $A_1$ is an increasing function of $c_1$ if $b_1>0$. If, on the other hand, $b_1<0$, then
$A_1$ becomes an increasing function of $c_1$ in the interval $0<c_1\leq{1\over 4b_1^2}$ and a decreasing function of $c_1$ in
the interval ${1\over 4b_1^2}<c_1<{1\over b_1^2}$. 
In the case of $b_1>0$, the amplitude $A_1$ remains finite in the limit of $a_1\rightarrow 0$ (i.e., infinite width) as opposed to the behavior of the usual soliton
for which the amplitude tends to zero in the limit of infinite width.
It then turns out that  the bright soliton $u_1$ from (4.4) reduces to an
algebraic soliton of the form 
$$u_1={{\rm i}c_1^{3\over 4}\,{\rm exp}\left[{\rm i}\left({1\over \sqrt{c_1}}x-\sqrt{c_1}t+\chi_{10}\right)\right]\over x+c_1t+x_0-{\rm i}{\sqrt{c_1}\over 2}},\qquad c_1={1\over b_1^2}, \eqno(4.7a)$$
and $|u_1|^2$ from (4.5) becomes
$$|u_1|^2={c_1^{3\over 2}\over (x+c_1t+x_0)^2+{c_1\over 4}}, \eqno(4.7b)$$
where we have put $\theta_{10}=a_1x_0-\delta_1$.  The similar peculiar structure of the solution has been found for the bright soliton solution of the DNLS equation [11]. \par
In the case of $b_1<0$, we can observe that  when the velocity lies in the interval ${1\over 4b_1^2}<c_1<{1\over b_1^2}$, the solution exhibits an interesting feature. 
Specifically, a large soliton propagates slower than a small soliton, as will be illustrated graphically in the interaction process of two bright solitons (see figure 3).
The amplitude-velocity relation described above is in striking contrast to that of the 1-soliton solution of the DNLS equation.
Actually, we see from (3.76) that  $A_1$ becomes
a decresing function of $|c_1|$ when $b_1<0$. \par
Last, it is instructive to compare the  solution (4.4) and (4.5) with that derived by the IST. 
We define the new parameters $\Delta_1$ and $\gamma_1$ by the relations
$a_1=\Delta_1^2\sin\,\gamma_1,\qquad b_1=-\Delta_1^2\cos\,\gamma_1, \qquad 0<\gamma_1<\pi$, 
and then replace $\theta_{10}$ by $\theta_{10}+\ln \Delta_1$ (or equivalently, $\theta_{1}$ by $\theta_{1}+\ln \Delta_1$). 
We find that  expressions (4.4) and (4.5) recast respectively to
$$u_1=-{1\over \Delta_1}\,{2{\rm i}\,{\rm e}^{\theta_1+{\rm i}\chi_1}\sin\,\gamma_1\over {\rm e}^{2\theta_1+{\rm i}\gamma_1} +1}, \eqno(4.8a)$$
$$|u_1|^2={1\over \Delta_1^2}\,{2\,\sin^2\gamma_1\over \cosh\,2\theta_1 +\cos\,\gamma_1},\eqno(4.8b)$$
which is just the 1-soliton solution presented in [3]. 
Note in these expressions that $A_1=2\,\sin{\gamma_1\over 2}/\Delta_1$ and $c_1=1/\Delta_1^2$. The algebraic soliton is generated from (4.8) 
if one puts $\gamma_1=\pi-\epsilon$ and
takes the limit $\epsilon\rightarrow +0$ [3]. 
\par
\medskip
\noindent{\it 4.2. Bright 2-soliton solution}\par
\medskip
\noindent The bright 2-soliton solution exhibits a variety of interesting features. Here, we investigate the interaction process of bright solitons
focusing on the asymptotic behavior of the solution.  As a result, we obtain the formulas for the phase shifts for each soliton.
 The corresponding  tau functions $f_2$ and $g_2$  are given by (3.1) and (3.2) with $N=2$.  Explicitly,
$$f_2=-{p_1^*p_2^*(p_1-p_2)(p_1^*-p_2^*)\over (p_1+p_1^*)(p_1+p_2^*)(p_2+p_1^*)(p_2+p_2^*)}
-{{\rm i}p_2^*\over (p_1+p_1^*)(p_2+p_2^*)}z_1z_1^*$$
$$-{{\rm i}p_1^*\over (p_1+p_1^*)(p_2+p_2^*)}z_2z_2^*
+{{\rm i}p_1^*\over (p_1+p_2^*)(p_2+p_1^*)}z_1z_2^*
+{{\rm i}p_2^*\over (p_1+p_2^*)(p_2+p_1^*)}z_2z_1^*$$
$$+{(p_1-p_2)(p_1^*-p_2^*)\over (p_1+p_1^*)(p_1+p_2^*)(p_2+p_1^*)(p_2+p_2^*)}z_1z_1^*z_2z_2^*, \eqno(4.9a)$$
$$g_2=-{{\rm i}p_2(p_1^*-p_2^*)\over p_1(p_2+p_1^*)(p_2+p_2^*)}z_1+{{\rm i}p_1(p_1^*-p_2^*)\over p_2(p_1+p_1^*)(p_1+p_2^*)}z_2
-{p_1^*(p_1-p_2)\over p_1p_2(p_1+p_1^*)(p_2+p_1^*)}z_1z_1^*z_2$$
$$+{p_2^*(p_1-p_2)\over p_1p_2(p_1+p_2^*)(p_2+p_2^*)}z_1z_2z_2^*. \eqno(4.9b)$$ \par
 Now, we order the magnitude of the velocity of each soliton in the $(x, t)$ coordinate system as $c_1>c_2$.
We first take the limit $t\rightarrow -\infty$  with $\theta_1$ being fixed. 
Since in this limit $|z_1|=$finite and $|z_2|\rightarrow \infty$, the leading-order asymptotics of $f_2$ and $g_2$ are found to be as
$$f_2 \sim {z_2z_2^*\over (p_1+p_1^*)(p_2+p_2^*)}\left\{{(p_1-p_2)(p_1^*-p_2^*)\over (p_1+p_2^*)(p_2+p_1^*)}z_1z_1^*-{\rm i}p_1^*\right\},\eqno(4.10a)$$
$$g_2 \sim {p_2^*(p_1-p_2)\over p_1p_2(p_1+p_2^*)(p_2+p_2^*)}z_1z_2z_2^*. \eqno(4.10b)$$
The asymptotic form of the 2-soliton solution $u_2$ evaluated by (4.10) is expressed as
$$u_2 \sim {p_1+p_1^*\over p_1}{z_1^\prime\over z_1^\prime{z_1^\prime}^*-{\rm i}p_1}, 
\qquad z_1^\prime=z_1\,{\exp}\left[-{\rm ln}\left\{{p_2(p_1+p_2^*)\over p_2^*(p_1-p_2)}\right\}\right]. \eqno(4.11)$$
Thus, the asymptotic  of $u_2$  takes the same profile as that of the 1-soliton solution (4.4) except the phase shifts. Specifically, 
$$u_2 \sim u_1(\theta_1+\Delta\theta_1^{(-)},\chi_1+\Delta\chi_1^{(-)}), \eqno(4.12a)$$
$$\Delta\theta_1^{(-)}=-\,{\rm ln}\left|{p_1+p_2^*\over p_1-p_2}\right|,
\qquad \Delta\chi_1^{(-)}=-\,{\rm arg}\,{p_1+p_2^*\over p_1-p_2}-{\rm arg}\,{p_2\over p_2^*}+\pi. \eqno(4.12b)$$ 
Next, we take the limit $t\rightarrow +\infty$  with $\theta_1$ being fixed. In this limit, $|z_1|=$finite and $|z_2|\rightarrow 0$.
The expressions corresponding to  (4.10) and (4.11) are given by
$$f_2 \sim -{{\rm i}p_2^*(p_1-p_2)(p_1^*-p_2^*)\over (p_1+p_1^*)(p_1+p_2^*)(p_2+p_1^*)(p_2+p_2^*)}\left\{{(p_1+p_2^*)(p_2+p_1^*)\over (p_1-p_2)(p_1^*-p_2^*)}\,z_1z_1^*-{\rm i}p_1^*\right\}, \eqno(4.13a)$$
$$g_2 \sim -{{\rm i}p_2(p_1^*-p_2^*)\over p_1(p_2+p_1^*)(p_2+p_2^*)}\,z_1, \eqno(4.13b)$$
$$u_2 \sim {p_1+p_1^*\over p_1}{z_1^{\prime\prime}\over z_1^{\prime\prime}{z_1^{\prime\prime}}^*-{\rm i}p_1}, 
\qquad z_1^{\prime\prime}=z_1\,{\exp}\left[{\rm ln}\left\{{p_2(p_1+p_2^*)\over p_2^*(p_1-p_2)}\right\}\right]. \eqno(4.14)$$
It follows from (4.14) that
$$u_2 \sim u_1(\theta_1+\Delta\theta_1^{(+)},\chi_1+\Delta\chi_1^{(+)}), \eqno(4.15a)$$
$$\Delta\theta_1^{(+)}=\,{\rm ln}\left|{p_1+p_2^*\over p_1-p_2}\right|,\qquad \Delta\chi_1^{(+)}=\,{\rm arg}\,{p_1+p_2^*\over p_1-p_2}+{\rm arg}\,{p_2\over p_2^*}+\pi.\eqno(4.15b)$$
In view of the fact that solitons propagate to the left and the trajectory of
the center position of the soliton is described by the equation $\theta_1+\Delta\theta_1^{(\pm)}=0$, we can  define the phase shifts 
$\Delta x_j=(\Delta\theta_j^{(+)}-\Delta\theta_j^{(-)})/a_j$ and $\Delta\chi_j=\Delta\chi_j^{(+)}-\Delta\chi_j^{(-)}$ which represent the total changes of the center
position and the phase of the $j$th  soliton $(j=1, 2)$, respectively.  Using (4.12) and (4.15), we then obtain
$$ \Delta x_1={2\over a_1}\,{\rm ln}\left|{p_1+p_2^*\over p_1-p_2}\right|,\qquad \Delta\chi_1=2\,{\rm arg}\,{p_1+p_2^*\over p_1-p_2}+2\,{\rm arg}\,{p_2\over p_2^*}.\eqno(4.16)$$
We can perform the similar asymptotic analysis while keeping $\theta_2$ fixed. Hence, we quote the final result. As $t\rightarrow -\infty$, the expression corresponding to (4.12) reads
$$u_2 \sim u_1(\theta_2+\Delta\theta_2^{(-)},\chi_2+\Delta\chi_2^{(-)}), \eqno(4.17a)$$
$$\Delta\theta_2^{(-)}=\,{\rm ln}\left|{p_2+p_1^*\over p_2-p_1}\right|,\qquad \Delta\chi_2^{(-)}=\,{\rm arg}\,{p_2+p_1^*\over p_2-p_1}+{\rm arg}\,{p_1\over p_1^*}+\pi,\eqno(4.17b)$$
whereas as $t\rightarrow +\infty$, the expression corresponding to (4.15) reads
$$u_2 \sim u_1(\theta_2+\Delta\theta_2^{(+)},\chi_2+\Delta\chi_2^{(+)}), \eqno(4.18a)$$
$$\Delta\theta_2^{(+)}=-\,{\rm ln}\left|{p_2+p_1^*\over p_2-p_1}\right|,\qquad \Delta\chi_2^{(+)}=-\,{\rm arg}\,{p_2+p_1^*\over p_2-p_1}-{\rm arg}\,{p_1\over p_1^*}+\pi.\eqno(4.18b)$$
\par
Thus, the total phase shifts are given by the formulas
$$ \Delta x_2=-{2\over a_2}\,{\rm ln}\left|{p_2+p_1^*\over p_2-p_1}\right|,\qquad \Delta\chi_2=-2\,{\rm arg}\,{p_2+p_1^*\over p_2-p_1}-2\,{\rm arg}\,{p_1\over p_1^*}.\eqno(4.19)$$
\par
Figure 2 shows the time evolution of a bright 2-soliton solution $U=|u|$ with positive $b_j\ (j=1, 2)$. See the solid line in fugure 1 which plots the amplitude-velocity relation for $b_1=1$.
In the present example, $A_1=1.99, A_2=1.44, c_1=0.99, c_2=0.20$ and the phase shifts are given by $\Delta x_1=2.00, \Delta x_2=-0.10$. The feature of the interaction process of solitons is similar to
that of the usual solitons, namely the large soliton propagates faster than the small soliton and the former one suffers a positive phase shift whereas  the latter
one suffers a negative phase shift. \par
\medskip
\begin{center}
\includegraphics[width=10cm]{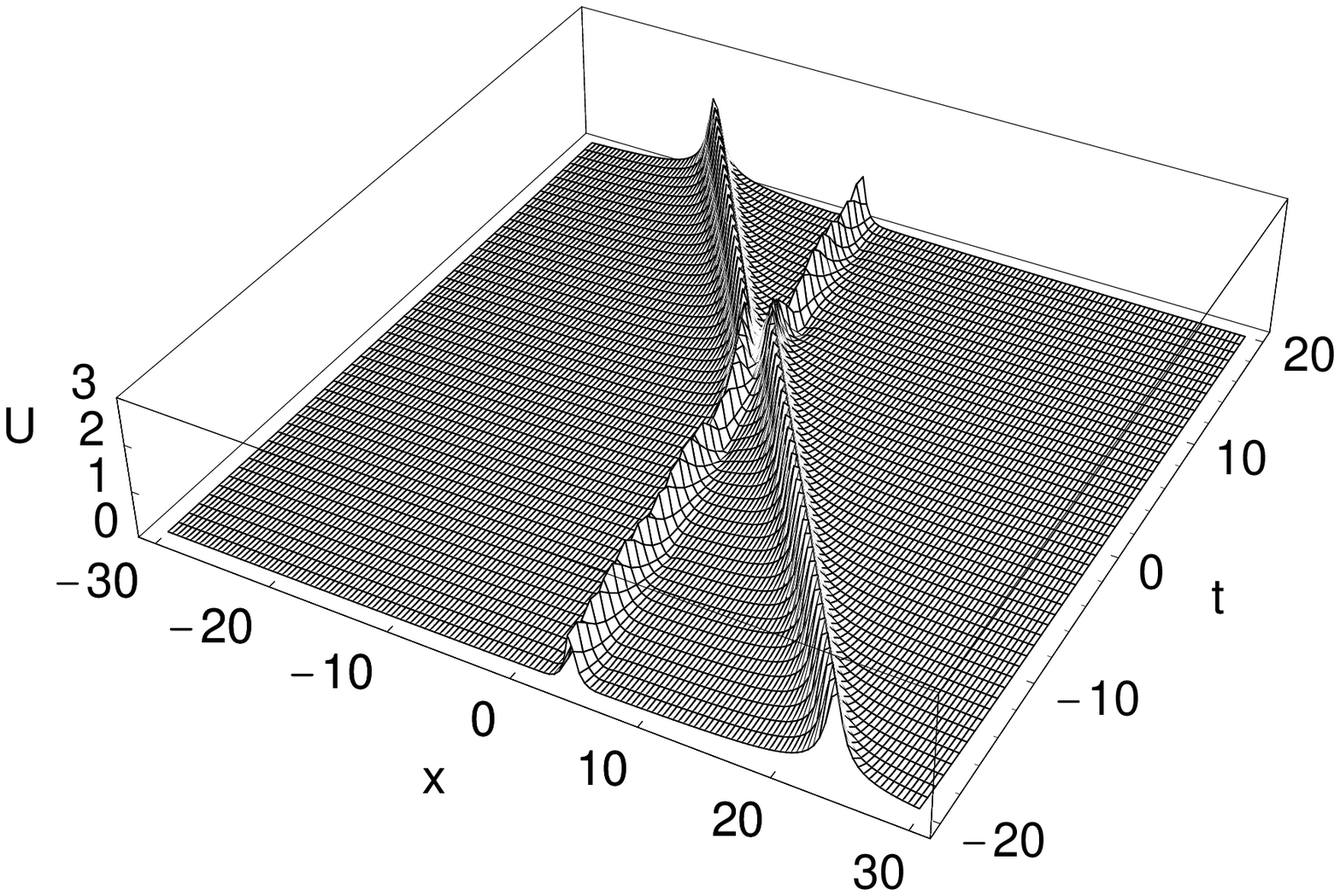}
\end{center}
\noindent {\bf Figure 2.} Time evolution of a bright 2-soliton solution with the parameters $a_1=0.1, a_2=2.0, b_1=1.0, b_2=1.0$. \par
\bigskip
Figure 3 depicts the similar plot for negative $b_j\ (j=1, 2)$ 
where $A_1=0.43, A_2=0.69, c_1=0.80, c_2=0.41$ and $\Delta x_1=3.55, \Delta x_2=-1.48$. 
\medskip
\begin{center}
\includegraphics[width=10cm]{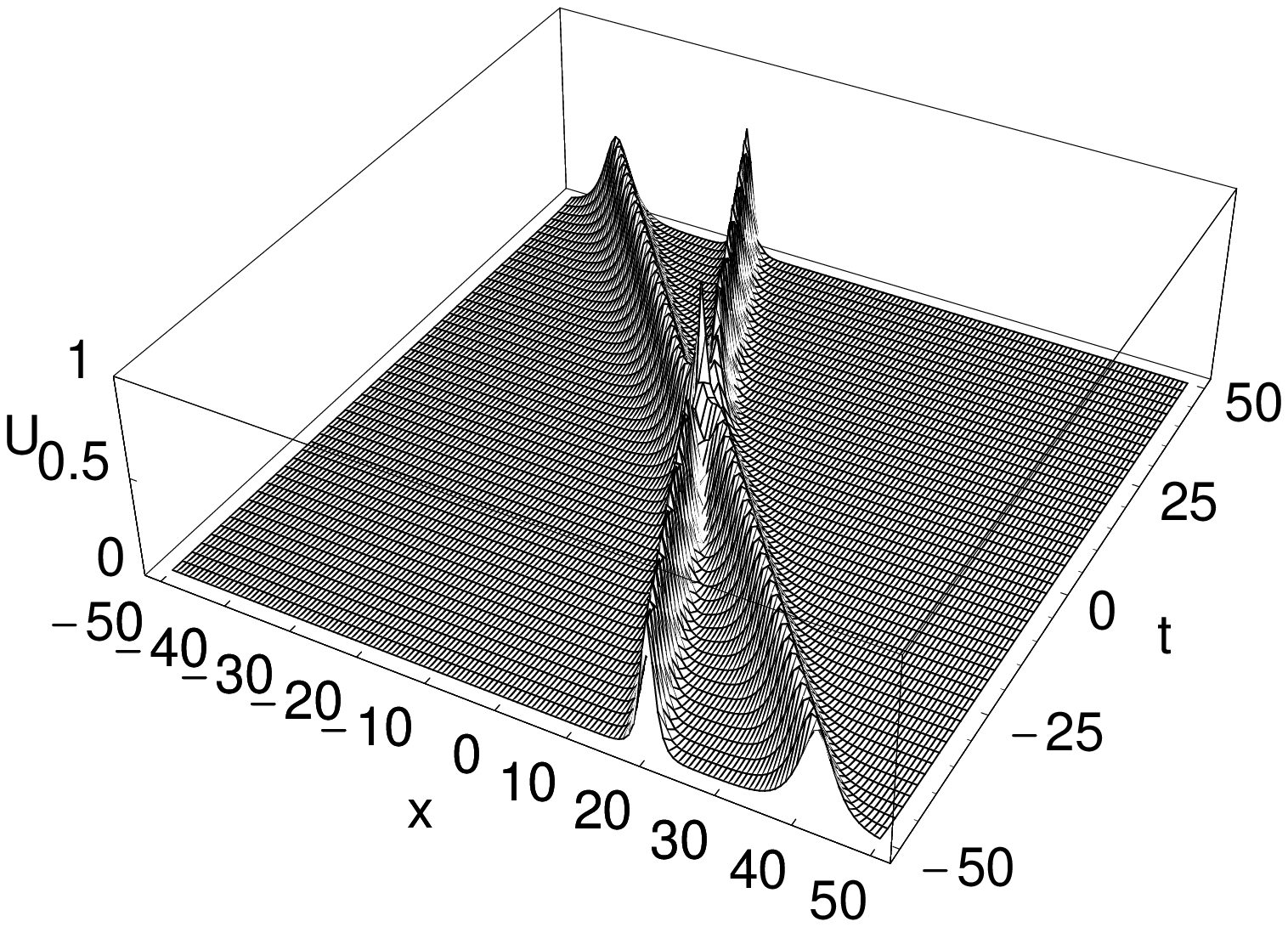}
\end{center}
\noindent {\bf Figure 3.} Time evolution of a bright 2-soliton solution with the parameters $a_1=0.5, a_2=1.2, b_1=-1.0, b_2=-1.0$. \par
\bigskip
In this case, the velocities $c_j (j=1, 2)$ are chosen to satisfy
the inequality ${1\over 4b_j^2}<c_j<{1\over b_j^2}$. Consequently, the velocity of each soliton becomes a decreasing function of the amplitude, as indicated by the broken
line in figure 1. Unlike the previous example, there appears a peculiar feature in the interaction process of solitons. Indeed, as evidenced from figure 3, the large soliton 
is seen to propagagate slower than the small soliton and suffers a negative phase shift. On the other hand, the small soliton suffers a positive phase shift. 
 \par
\bigskip
\noindent{\it 4.3. Bright $N$-soliton solution}\par
\medskip
\noindent 
The asymptotic analysis for the bright $N$-soliton solution can be performed  straightforwardly following the 2-soliton case. 
Hence, without entering into the  detailed calculation,  we describe the main results only. 
To proceed, let us order the magnitude of the velocity of each soliton  as $c_1>c_2> ... >c_N$. We take the limit $t \rightarrow -\infty$ with $\theta_n$ being finite. 
Since in this limit $|z_j|\rightarrow 0$ for $j<n$ and $|z_j|\rightarrow \infty$ for $n<j$, we find that the leading-order asymptotic of
the tau function $f=f_N$  from (3.1) and (3.2a)  can be written in the form
$$f_N \sim d(1, 2, ..., n)d(n+1, n+2, ..., N)\prod_{k=1}^{n-1}(-{\rm i}p_k)\prod_{j=n+1}^N(z_jz_j^*)\,(d_nz_zz_n^*-{\rm i}p_n^*), \eqno(4.20a)$$
where
$$d(m, m+1, ..., n)=\left|\left({1\over p_j+p_k^*}\right)_{m\leq j,k\leq n}\right|={\prod_{m\leq j<k\leq n}(p_j-p_k)(p_j^*-p_k^*)\over \prod_{m\leq j,k\leq n}(p_j+p_k^*)}, \quad 1\leq m<n\leq N, \eqno(4.20b)$$
is the Cauchy determinant and
$$d_n={d(1, 2, ..., n-1)d(n, n+1, ..., N)\over d(1, 2, ..., n)d(n+1, n+2, ..., N)}
={\prod_{j=n+1}^N{(p_n-p_j)(p_n^*-p_j^*)\over (p_n+p_j^*)(p_n^*+p_j)}\over \prod_{j=1}^{n-1}{(p_n-p_j)(p_n^*-p_j^*)\over (p_n+p_j^*)(p_n^*+p_j)}}. \eqno(4.20c)$$
Similarly, $g_N$ has the asymptotic form
$$g_N \sim {(-1)^N\over p_n\,}d(1, 2, ..., n-1)d(n+1, n+2, ..., N)$$
$$\times \prod_{k=1}^{n-1}(-{\rm i}p_k^*)
\prod_{j=1}^{n-1}{p_j(p_n^*-p_j^*)\over p_j^*(p_n^*+p_j)}
\prod_{k=n+1}^N{p_k^*(p_n-p_k)\over p_k(p_n+p_k^*)}
\prod_{j=n+1}^N(z_jz_j^*)\,z_n. \eqno(4.21)$$
Taking into account  the relation
$$d(1, 2, ..., n)={1\over p_n+p_n^*}\prod_{j=1}^{n-1}{(p_n-p_j)(p_n^*-p_j^*)\over (p_n+p_j^*)(p_n^*+p_j)}\,d(1, 2, ..., n-1), \eqno(4.22)$$
we obtain from (4.20) and (4.21) the asymptotic form of the $N$-soliton solution
$$u_N \sim (-1)^N{p_n+p_n^*\over p_n}\prod_{j=1}^{n-1}{p_j(p_n+p_j^*)\over p_j^*(p_n-p_j)}\prod_{k=n+1}^N{p_k^*(p_n-p_k)\over p_k(p_n+p_k^*)}\,{z_n\over d_nz_nz_n^*-{\rm i}p_n^*}. \eqno(4.23)$$
This expression can be rewritten in terms of the 1-soliton solution as
$$u_N \sim u_1(\theta_n+\Delta\theta_n^{(-)}, \chi_n+\Delta\chi_n^{(-)}), \eqno(4.24a)$$
with
$$\Delta\theta_n^{(-)}=\sum_{j=1}^{n-1}{\rm ln}\left|{p_n+p_j^*\over p_n-p_j}\right|-\sum_{j=n+1}^{N}{\rm ln}\left|{p_n+p_j^*\over p_n-p_j}\right|, \eqno(4.24b)$$
$$\Delta\chi_n^{(-)}=\sum_{j=1}^{n-1}\left\{{\rm arg}\left({p_n+p_j^*\over p_n-p_j}\right)+{\rm arg}\left({p_j\over p_j^*}\right)\right\}
-\sum_{j=n+1}^{N}\left\{{\rm arg}\left({p_n+p_j^*\over p_n-p_j}\right)+{\rm arg}\left({p_j\over p_j^*}\right)\right\}+N\pi. \eqno(4.24c)$$
 \par
 As $t \rightarrow +\infty$, on the other hand, the asymptotic form of $u_N$ is found to be as
$$u_N \sim (-1)^N{p_n+p_n^*\over p_n}\prod_{j=n+1}^{N}{p_j(p_n+p_j^*)\over p_j^*(p_n-p_j)}\prod_{k=1}^{n-1}{p_k^*(p_n-p_k)\over p_k(p_n+p_k^*)}\,{z_n\over d_n^{-1}z_nz_n^*-{\rm i}p_n^*}, \eqno(4.25)$$
which leads to the expression 
$$u_N \sim u_1(\theta_n+\Delta\theta_n^{(+)}, \chi_n+\Delta\chi_n^{(+)}), \eqno(4.26a)$$
with
$$\Delta\theta_n^{(+)}=-\Delta\theta_n^{(-)}, \qquad \Delta\chi_n^{(+)}=-\Delta\chi_n^{(-)}+2N\pi. \eqno(4.26b)$$
\par
We see from (4.24) and (4.26) that in the rest frame of reference, the asymptotic form of the bright $N$-soliton solution can be represented by a superposition of $N$ independent
bright 1-soliton solutions, the only difference being the phase shifts.
It follows from (4.24) and (4.26) that the formulas for the total phase shifts of the $n$th soliton are given by 
$$\Delta x_n={2\over a_n}\left\{\sum_{j=n+1}^{N}{\rm ln}\left|{p_n+p_j^*\over p_n-p_j}\right|-\sum_{j=1}^{n-1}{\rm ln}\left|{p_n+p_j^*\over p_n-p_j}\right|\right\}, \eqno(4.27a)$$
$$\Delta\chi_n=2\sum_{j=n+1}^{N}\left\{{\rm arg}\left({p_n+p_j^*\over p_n-p_j}\right)+{\rm arg}\left({p_j\over p_j^*}\right)\right\}
-2\sum_{j=1}^{n-1}\left\{{\rm arg}\left({p_n+p_j^*\over p_n-p_j}\right)+{\rm arg}\left({p_j\over p_j^*}\right)\right\}. \eqno(4.27b)$$
\par
The above formulas reduce to (4.16) and (4.19) for the special case of $N=2$. They clearly show that each soliton has pairwise interactions with
other solitons, namely there are no many-particle collisions among solitons.
\par
\bigskip
\leftline{\bf 5. Concluding remarks} \par
\bigskip
\noindent In this paper, we have presented two different expressions of the bright $N$-soliton solution of the FL DNLS equation.
The exact method of solution developed here is purely algebraic which does not recourse to the IST.
The $N$-soliton solution given by theorem 3.1 has been obtained by the dressing method [4] while the form presented by theorem 3.2 is new.
The equivalence of both expressions has been demonstrated by a straightforward computation  using the properties of the Cauchy matrix. 
We have also shown that the bright $N$-soliton solution of the DNLS equation
can be  derived from that of the FL DNLS equation through a simple relation. This fact has been stemmed from the analysis of a trilinear equation
among the tau functions. Last, we have investigated in detail the properties of the solutions and found some new features. \par
In conclusion, we briefly comment on the dark soliton solutions. The construction of soliton solutions under nonvanishing boundary conditions 
is difficult to perform when compared with that under vanishing
boundary conditions.  Specifically, the analysis by means of the IST requires the delicate discussion on the spectral problem. 
On the other hand, the bilinear transformation method is applicable easily to obtain  soliton solutions. In part II of the present study, we will
address the construction of the dark $N$-soliton solution of the FL DNLS equation on a  background of a plane wave. 
The system of bilinear equations for the equation is almost the same as that of the bright soliton case.
However, a difficulty arises due to  the constraint imposed on the complex parameters $p_j\ (j=1, 2, ..., N)$, as in the case of the similar problem for the 
NLS equation  in which  $p_j$ lie on a circle  in the complex plane [12].   In particular, this fact must be  used explicitly to
prove the bilinear equation corresponding to equation (2.2). We will show that the trilinear equation analogous to (2.10) play an important role in the proof and the inspection of
the equation also provides the dark $N$-soliton solution of the DNLS equation. \par

\bigskip
\leftline{\bf Acknowledgement}\par
\bigskip
\noindent This work was partially supported by the Grant-in-Aid for Scientific Research (C) No. 22540228 from Japan Society for the Promotion of Science. \par

\newpage

\leftline{\bf References}\par
\begin{enumerate}[{[1]}]
\item Fokas A S 1995 {On a class of physically important integrable equations} {\it Physica D} {\bf 87} 145-150
\item Lenells J 2009 {Exactly solvable model for nonlinear pulse propagation in optical fibers} {\it Stud. Appl. Math.} {\bf 123} 215-232
\item Lenells J and Fokas A S 2009 {On a novel integrable generalization of the nonlinear Schr\"odinger equation} {\it Nonlinearity} {\bf 22} 11-27
\item Lenells J 2010 {Dressing for a integrable generalization of the nonlinear Schr\"odinger equation} {\it J. Nonlinear Sci.} {\bf 20} 709-722
\item Matsuno Y 1984 {\it Bilinear Transformation Method} (New York: Academic)
\item Hirota R 2004 {\it The Direct Method in Soliton Theory} (New York: Cambridge)
\item Vein R  and Dale P 1999 {\it Determinants and Their Applications in Mathematical Physics} (New York: Springer)
\item Huang N N and Chen Z Y 1990 {Alfven solitons} {\it J. Phys. A: Math. Gen.} {\bf 23} 439-453
\item Matsuno Y 2011 {The {\it N}-soliton solution of a two-component modified nonlinear Schr\"odinger equation} {\it Phys. Lett.} {\bf A 375} 3090-3094
\item Matsuno Y 2011 {The bright {\it N}-soliton solution of a multi-component modified nonlinear Schr\"odinger equation} {\it J. Phys. A: Mat. Theor.} {\bf 44} 495202
\item Kaup D J and Newell A C 1977 {An exact solution of a derivative nonlinear Schr\"odinger equation} {\it J. Math. Phys.} {\bf 19} 798-801
\item Faddeev L D and Takhtajan L A 1987  {\it Hamiltonian Methods in the Theory of Solitons} (Berlin: Springer)
\end{enumerate}

\end{document}